\documentclass[journal=nalefd,manuscript=letter]{achemso}
\usepackage[version=3]{mhchem}
\usepackage[separate-uncertainty=true]{siunitx}
\usepackage{xcolor}
\newcommand{\figurescale}{1}
\newcommand{\onecolfigurescale}{1.5}

\author{M.~Blauth}
\affiliation[WSI]
{Walter Schottky Institut and Physik Department, Technische Universit\"at M\"unchen, Am Coulombwall 4, 85748 Garching, Germany}
\alsoaffiliation{Nanosystems Initiative Munich (NIM), Schellingstr. 4, 80799 Munich, Germany}
\author{M.~J\"urgensen}
\author{G.~Vest}
\author{O.~Hartwig}
\author{M. Prechtl}
\affiliation[WSI]
{Walter Schottky Institut and Physik Department, Technische Universit\"at M\"unchen, Am Coulombwall 4, 85748 Garching, Germany}
\author{J.~Cerne}
\affiliation[WSI]
{Walter Schottky Institut and Physik Department, Technische Universit\"at M\"unchen, Am Coulombwall 4, 85748 Garching, Germany}
\alsoaffiliation{Department of Physics, University at Buffalo, The State University of New York, Buffalo, New York 14260, USA}
\author{\newline J.~J.~Finley}
\affiliation[WSI]
{Walter Schottky Institut and Physik Department, Technische Universit\"at M\"unchen, Am Coulombwall 4, 85748 Garching, Germany}
\alsoaffiliation{Nanosystems Initiative Munich (NIM), Schellingstr. 4, 80799 Munich, Germany}
\email{finley@wsi.tum.de}
\author{M.~Kaniber}
\email{kaniber@wsi.tum.de}
\affiliation[WSI]
{Walter Schottky Institut and Physik Department, Technische Universit\"at M\"unchen, Am Coulombwall 4, 85748 Garching, Germany}
\alsoaffiliation{Nanosystems Initiative Munich (NIM), Schellingstr. 4, 80799 Munich, Germany}

\title{Coupling single photons from discrete quantum emitters in WSe$_2$ to lithographically defined plasmonic slot-waveguides}

\keywords{Plasmonics, Quantum plasmonics, Localized excitons, WSe$_2$, Slot waveguide}

\begin{document}

\begin{abstract}
We report the observation of the generation and routing of single plasmons generated by localized excitons in a WSe$_2$ monolayer flake exfoliated onto lithographically defined Au-plasmonic waveguides.
Statistical analysis of the position of different quantum emitters shows that they are $(3.3 \pm 0.7) \times$ more likely to form close to the edges of the plasmonic waveguides.
By characterizing individual emitters we confirm their single-photon character via the observation of antibunching of the signal ($g^{(2)}(0) = 0.42$) and demonstrate that specific emitters couple to the modes of the proximal plasmonic waveguide.
Time-resolved measurements performed on emitters close to, and far away from the plasmonic nanostructures indicate that Purcell factors up to $15 \pm 3$ occur, depending on the precise location of the quantum emitter relative to the tightly confined plasmonic mode.
Measurement of the point spread function of five quantum emitters relative to the waveguide with
\SI{<50}{\nano\meter} precision are compared with numerical simulations to demonstrate potential for higher increases of the coupling efficiency for ideally positioned emitters.
The integration of such strain-induced quantum emitters with deterministic plasmonic routing is a step toward deep-subwavelength on-chip single quantum light sources.
\end{abstract}

\section{Introduction}
Downscaling of integrated devices for information technologies is fueled by the need to reduce the energy overhead per bit of data processed.\cite{Schaller.1997,Ha.,Wu.,Kish.2002}
It was already recognized several decades ago that shifting from electronic to photonic devices\cite{Ozbay.2006,Sorger.2012} promises ultra high-rate data processing, with maximum accessible clock speeds beyond $\sim$THz\cite{Heuring.1992}.
In terms of the \textit{energy} required to process a \textit{single bit} of information, all-optical approaches lead the way. Non-linear interactions can occur in nano-photonic devices and circuits already at the few-photon limit, corresponding to an energy-per-bit budget in the deep sub-fJ regime \cite{Nozaki.2010}.
As such, research into quantum light sources capable of delivering non-classical states of light (single and few-photon states)\cite{Aharonovich.2016} into integrated photonic circuits are of strong interest, especially if they are capable of operating at elevated temperature.
In these respects, transition metal dichalcogenides (TMDCs) have captured the attention of many groups worldwide.
Monolayers of 2H-stacked TMDCs are direct gap semiconductors\cite{Splendiani.2010,Mak.2010} and have very large exciton binding energies (\SIrange{\sim 200}{500}{\milli\electronvolt}) and low excitonic Bohr radii of only a few nanometers\cite{Ugeda.2014,Chernikov.2014}.
Moreover, they can exhibit near-unity internal quantum efficiencies when suitably processed\cite{Amani.2015} and the local exciton binding energy is sensitive to the proximal dielectric environment at the nanometer scale\cite{Roesner.2016}, and the presence of strain\cite{Tonndorf.2015}.
It has been shown that single photon emitters occur naturally in mechanically exfoliated WSe$_2$ \cite{Chakraborty.2015,He.2015,Koperski.2015,Srivastava.2015,Tonndorf.2015} and that they can be positioned by engineering of the local strain field\cite{Kern.2016,PalaciosBerraquero.2017,Branny.}.
At the same time, the very strong spin-orbit interactions in TMDCs provide unique optical access to spin and valley degrees of freedom \cite{DiXiao.2012,Cao.2012,Mak.2012,Zeng.2012,Sallen.2012} providing additional scope for encoding information.
Beyond low-energy switching, a clear \textit{disadvantage} of integrated photonic approaches to information processing is that the lower bound on the size of conventional components are fundamentally limited to the order of the optical wavelength\cite{Born.2016}.
This results in far lower integration densities as compared to integrated electronics.
In this respect, plasmonics offers one way to deliver deep-subwavelength confinement at optical frequencies \cite{Ozbay.2006} and, when combined with novel light-emitting materials, this raises the potential for photonic and quantum devices at the nanoscale.
Recent experiments have demonstrated that both free exciton \cite{Goodfellow.2014,Lee.2015,Zhu.2016} and localized exciton \cite{Cai.2017} emission can be coupled to plasmonic modes in chemically synthesized nanowires and dielectric waveguides\cite{Tonndorf.2017}.

Here, we mechanically exfoliate a monolayer flake of WSe$_2$ and transfer it onto a lithographically defined plasmonic waveguide.
This enables us to probe interactions between localized excitons in the WSe$_2$ flake and tightly confined plasmonic modes.
The use of electron-beam lithography provides full control over the position and geometry of the plasmonic waveguide, facilitating deterministic routing of single photons (and plasmons) on-chip.
The lithographically defined waveguide creates a non-planar substrate topography that we show results in local strain induced, discrete emitters in the monolayer.
The occurrence of such quantum emitters is shown to be $3.3 \pm 0.7\times$ more likely in the immediate vicinity ($\leq$500 nm) of the waveguide ends, as compared to unpatterned regions of the sample.
The quantum nature of the emitters is confirmed by measuring the second order intensity correlation function and spatially resolved measurements demonstrate that single photons are selectively coupled to the plasmonic waveguide mode.
Using time-resolved spectroscopy we show that emitters close to the waveguide (\SI{< 500}{\nano\meter}) exhibit Purcell factors in the range $F_P\sim2-15\times$ and, by carefully determining the position of five quantum emitters relative to the waveguide with sub \SI{50}{\nano\meter} precision via their point spread function, and performing numerical simulations we demonstrate the potential for significant further increases in coupling efficiency.
Our results pave the way towards novel on-chip single plasmon light sources at the nanoscale with the possibility for integration.

\section{Results and discussion}
\begin{figure}[!ht]
\scalebox{\figurescale}{\includegraphics{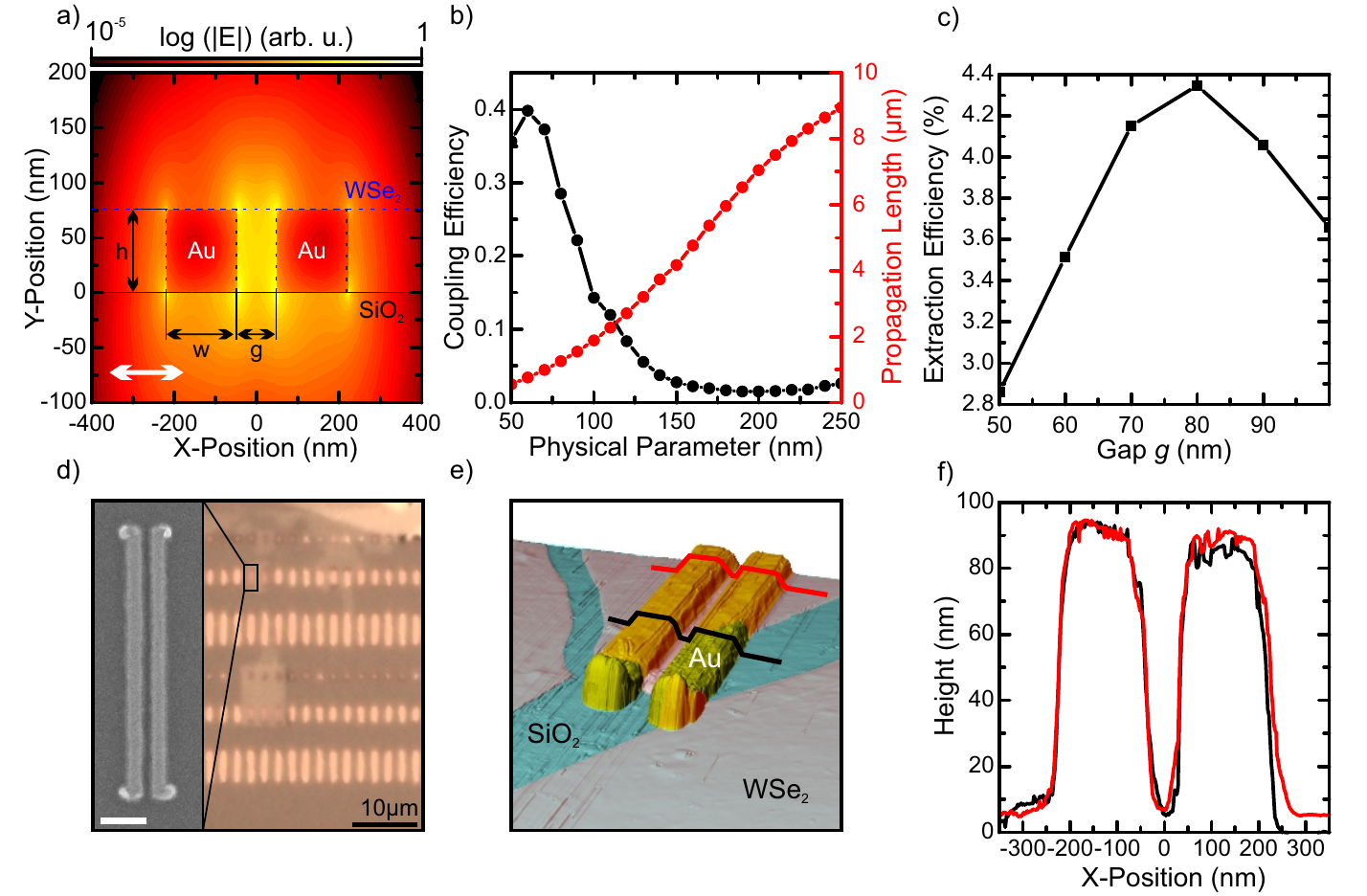}}
\renewcommand{\figurename}{Figure}
\caption{\label{fig1}
Overview of the composite structure.
(a) Electric field distribution of the antisymmetric plasmonic mode supported by the slot waveguide system calculated using Lumerical MODE Solution \cite{LumericalSolutionsInc..}. Labels indicate geometric parameters gap $g$, width $w$ and height $h$. White arrow indicates linear polarization orthogonal to the waveguide axis.
(b) Geometry dependent trade-off between induced light-matter interaction and propagation length for $g = w = h$.
(c) Combined figure of merit for light extraction from the TMDC monolayer.
(d) Optical microscope image of the fabricated plasmonic slot waveguide array covered by a WSe$_2$ monolayer flake. Inset: SEM image of an individual waveguide including outcoupling structures at both ends, scale bar: \SI{0.5}{\micro\meter}.
(e) False-color, perspective view of an atomic force microscope (AFM) image of the combined system. Labels indicate SiO$_2$ substrate (light blue), WSe$_2$ monolayer (grey) and slot waveguide (yellow), respectively. Curves (black, red) indicate positions of height profiles displayed in panel (f).
(f) AFM profiles of fully covered waveguide location (red) and partially covered waveguide (black).
}
\end{figure}

The plasmonic slot waveguides investigated consist of two metal strips separated by a dielectric slot.
Figure~\ref{fig1}(a) depicts a cross-sectional sketch for the composite system consisting of a  WSe$_2$ monolayer (blue dotted line) covering two metal bars with height $h$ and width $w$ separated by a gap $g$ on top of a SiO$_2$ substrate.
The false color scale displays the electric field distribution of the antisymmetric fundamental mode for $w = \SI{172}{\nano\meter}$, $g = \SI{96}{\nano\meter}$ and $h = \SI{75}{\nano\meter}$ computed using a finite difference eigenmode solver \cite{LumericalSolutionsInc..}.
For this mode, the electric field is maximum at the inner edges of the plasmonic waveguide and the plasmonic field is polarized along the x-axis as indicated by the white arrow.
Figure~\ref{fig1}(b) shows the coupling efficiency, i.e., the probability of an exciton decaying into the plasmonic modes supplied by the waveguide, as a function of scaling $g=w=h$.
Based on the dipole interaction Hamiltonian, the coupling efficiency scales with $|E_{in-plane}|^2$, thus, for small feature sizes, strong light-matter interaction is anticipated due to the high local field enhancement.
The coupling efficiency decreases with increasing structure size, becoming negligibly small once $g=w=h=\SI{150}{\nano\meter}$ due to weak confinement of the plasmonic mode.
Conversely, the propagation length (red curve) increases with increasing physical dimensions of the structure.
We optimize the geometry of our slot-waveguide structures to maximize the product of the coupling efficiency and waveguide transmission for a fixed design length of the waveguides of \SI{3}{\micro\meter}, as shown in Fig.~\ref{fig1}(c) as a function of gap $g$.
As the extraction efficiency does not vary strongly with the gap width, we conclude that this geometry is robust with respect to fabrication deviations.
More information about the scaling of the geometrical parameters can be found in the Supplementary Materials.

Slot waveguide arrays were fabricated on a SiO$_2$ substrate using electron beam lithography and gold evaporation.  Here, we use $g = \SI{96}{\nano\meter}$, $w = \SI{172}{\nano\meter}$, $h = \SI{75}{\nano\meter}$ and lengths of $\SI{1}{\micro\meter}$, $\SI{3}{\micro\meter}$ and $\SI{6}{\micro\meter}$ optimized for extraction efficiency.
In a following step, an all-dry transfer of monolayer WSe$_2$ \cite{CastellanosGomez.2014} is performed to cover large parts of the waveguide array.
Figure~\ref{fig1}(d) shows an optical microscope image of an array of plasmonic slot waveguides fully covered by a WSe$_2$ monolayer.
The scanning electron microscopy (SEM) image in the inset shows a detailed view of a single slot waveguide including outcoupling structures at both ends optimized for enhanced far-field coupling.

Figure~\ref{fig1}(e) depicts an atomic force microscope (AFM) image of an individual plasmonic waveguide (yellow), partially covered by a WSe$_2$ monolayer (grey).
Red and black curves indicate the locations of two height profiles visible in Fig.~\ref{fig1}(f).
For the left metal slab, both profiles are in good agreement, whereas the right slab is covered and uncovered for the red and the black profile, respectively, resulting in a different edge steepness of the rightmost edge.
These observations indicate close adhesion of the monolayer to the underlying metal structures and lead to the expectation of increased tensile strain in the flake close to the waveguide edges.

\begin{figure}[!ht]
\scalebox{\onecolfigurescale}{\includegraphics{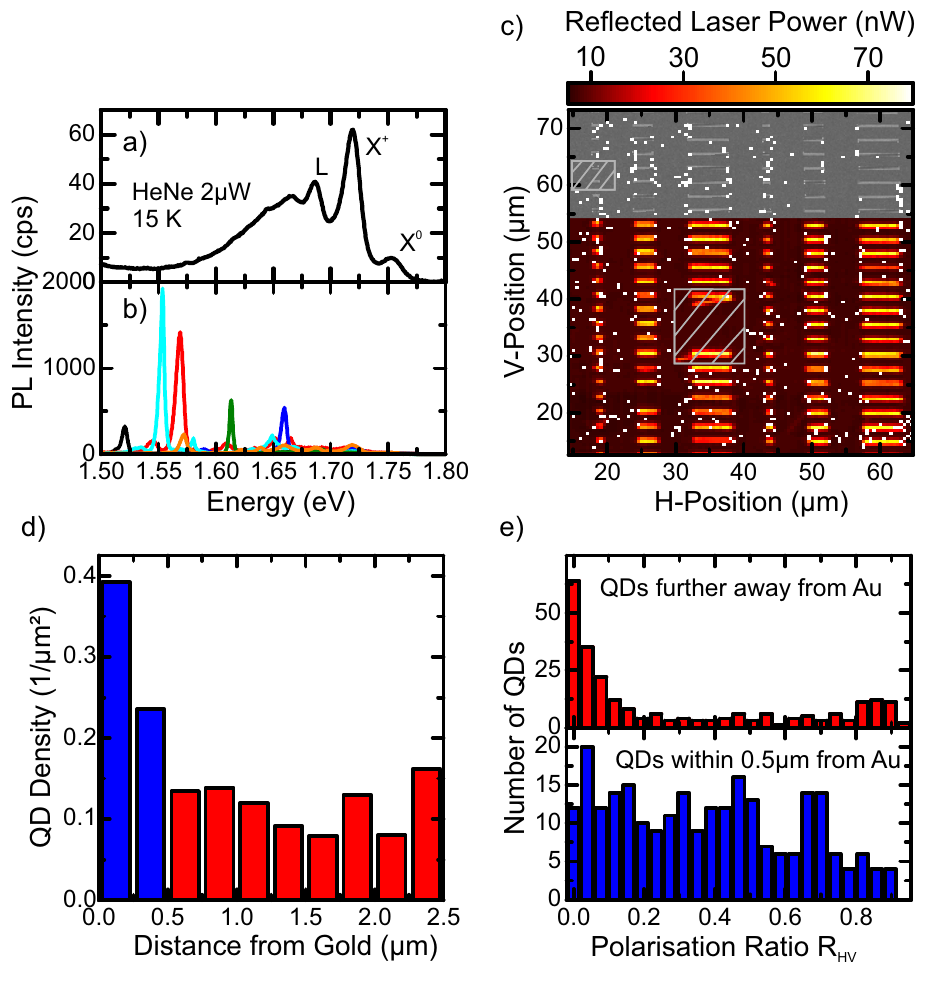}}
\renewcommand{\figurename}{Figure}
\caption{\label{fig2}
Statistical analysis of sharp emission lines occurring at the composite flake-waveguide system.
(a) Spatially averaged photoluminescence (PL) spectrum of large WSe$_2$ monolayer region, labels indicating the respective known spectral features.
(b) Selected PL spectra featuring sharp emission lines at different positions strongly deviating from the average spectrum.
(c) Large area confocal PL scan of sample region exhibiting plasmonic waveguides. Reflected laser power (red color) indicates the positions of the individual waveguides. Superimposed top part shows precise correlation of recorded power data and high-resolution SEM image. White dots indicate locations with isolated sharp emission lines comparable to panel (b). Shaded rectangles indicate regions ignored in analysis due to fabrication deviations.
(d) QD density as a function of distance of the emitter from the closest waveguide edge exhibiting significantly enhanced QD density for distances below \SI{0.5}{\micro\meter} (blue).
(e) Histogram of number of QDs as a function of degree of polarization measured in H-V basis for QDs closer (farther away) than \SI{0.5}{\micro\meter} in blue (red).
}
\end{figure}
To determine the influence of the plasmonic waveguides on the TMDC monolayer, we confocally recorded PL from a ca. \SI{3000}{\micro\meter\squared} sized region of the samples at cryogenic temperatures (\SI{15}{\kelvin}) using HeNe excitation (\SI{1.96}{\electronvolt}) with a \SI{2}{\micro\watt} excitation power.
Figure~\ref{fig2}(a) depicts the spectrally resolved PL intensity averaged over a large region of the sample consisting of both, a pristine WSe$_2$ monolayer and the combined system of plasmonic slot waveguides covered by a WSe$_2$ monolayer.
As reported in literature \cite{Jones.2013}, we observe the neutral (X$^0$), charged (X$^+$) and localized (L) emission peaks.
Moreover, a broad low-energy tail below \SI{1.67}{\electronvolt} is visible.
Figure~\ref{fig2}(b) presents a selection of individual PL spectra, revealing spatially strongly localized emission with linewidths between \SI{2}{\milli\electronvolt} and \SI{20}{\milli\electronvolt} distributed in energy between \SI{1.5}{\electronvolt} and \SI{1.67}{\electronvolt}, thereafter referred to as quantum dots (QD).
To study the spatial distribution of these QD-like emission lines we simultaneously recorded spatially resolved PL intensity and the reflection of the excitation laser, the latter being displayed in Fig.~\ref{fig2}(c).
Due to the enhanced reflectivity of the gold plasmonic waveguides, their position is determined by the reflected excitation laser power.
Thus, the positions of the individual waveguides could be extracted with high precision clearly reflecting the arrangement defined during fabrication.
To visualize this agreement, the upper part of Fig.~\ref{fig2}(c) shows an SEM image of this sample location superimposed onto the reflected laser topography image.
All recorded spectra were individually analyzed for sharp emission lines (for details on selection criteria, see Supplementary Materials) and positions featuring at least one sharp emission line are marked by a white pixel on Fig.~\ref{fig2}(c). (Shaded rectangles depict regions deviating significantly from the sample design that are ignored to avoid analysis artifacts from fabrication imperfections.)
Figure~\ref{fig2}(d) shows a quantitative analysis of the density of sharp emission lines as a function of their respective distance to the closest gold edge.
For distances shorter than \SI{0.5}{\micro\meter}, indicated by blue bars of the figure, the density of quantum emitters is found to be significantly enhanced compared to larger distances, shown in red. At its highest value of \SI{0.39}{\per\micro\meter\squared} the emitter density at the gold edges is enhanced by a factor of $(3.3 \pm 0.7)\times$ with respect to the average density of \SI{0.11 \pm 0.03}{\per\micro\meter\squared} at distances larger than \SI{0.5}{\micro\meter}.
This enhanced emitter density close to gold structures indicates that the formation probability is closely related to the substrate topography, in good agreement with the findings of previous studies\cite{Kern.2016,Branny.,PalaciosBerraquero.2017}.
Reference to Fig.~\ref{fig2}(c) also shows that the positions at the waveguide ends exhibit a locally enhanced emitter density which we attribute to the formation of a two-dimensional strain profile, whereas along the waveguide, the strain profile is mainly dominated by one-dimensional strain reducing the probability for full exciton confinement (see Supplementary Materials).
Therefore, the observed emitter density enhancement presented underestimates the structural influence of the waveguide, providing a lower bound.

When performing confocal PL measurements, we simultaneously recorded polarization-resolved spectra with the detection polarization along the waveguide axis (H-polarization), and orthogonal to the axis (V-polarization).  Thus, we define a polarization ratio $R_{HV} = |\frac{I_V - I_H}{I_V + I_H}|$ for each measured discrete emitter in H-V basis.
Figure~\ref{fig2}(d) shows a histogram of the distribution of measured polarization ratios.
The top panel in red shows the distribution of polarization ratio for emitters located $\geq$\SI{0.5}{\micro\meter} away from the closest gold edge, corresponding to the data set indicated in red in panel (c).
The distribution shows a clear maximum around zero and is significantly reduced for higher polarization ratios indicating that emitters forming far from plasmonic waveguides are predominantly linearly unpolarized.
The bottom panel of Fig.~\ref{fig2}(d) depicts the polarization ratio distribution for emitters close to the gold structures.
Conversely, the distribution of polarization ratio is found to decrease slowly for increasing polarization ratio revealing that the polarization is enhanced for emitters \textit{close} to the plasmonic structures with respect to their unperturbed counterparts.
This is consistent with the linear polarization supported by the plasmonic modes, thus, emitters coupling to these modes are expected to reflect this polarization.
In addition, both directions H and V coincide with the edges of the plasmonic waveguides and as known from literature \cite{Kern.2016}, the emission polarization is defined by the external strain fields.
Both contributions underpin the role of the waveguide in the formation of proximal localized emitters.

\begin{figure}[!ht]
\scalebox{\onecolfigurescale}{\includegraphics{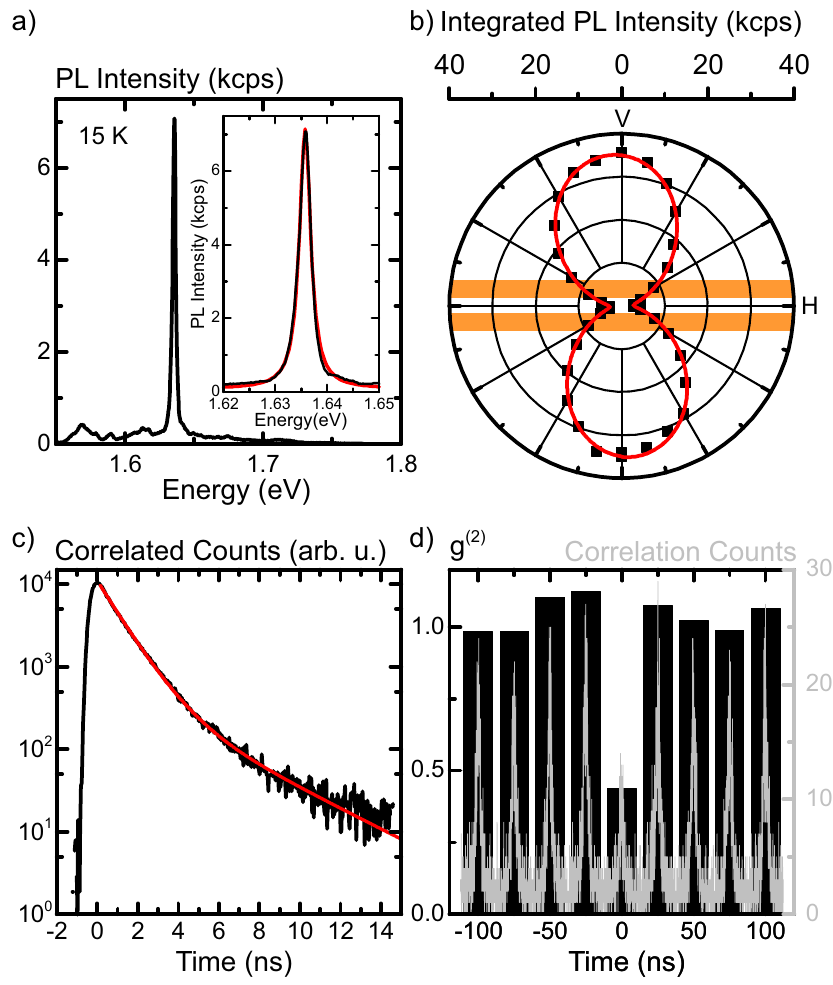}}
\renewcommand{\figurename}{Figure}
\caption{\label{fig3}
Characterization of an individual emitter at the waveguide end.
(a) Typical spectrum of a single sharp emission line. Inset: Zoomed spectrum including Lorentzian fit (red) to the raw data (black). (b) Detection polarization measurement of previous emitter (black data points) and sin$^2$ fit (red) revealing a degree of linear polarization of \SI{85.7}{\percent} orthogonal to the waveguide axis (orange sketch).
(c) Time resolved PL measurement indicating QD lifetime of \SI{1.07 \pm 0.01}{\nano\second} and corresponding bi-exponential fit (red).
(d) Second-order correlation function obtained from confocal PL measurement in pulsed excitation. Black bars indicate signal binned to the repetition frequency of \SI{40}{\mega\hertz} yielding $g^{(2)}(0) = \SI{0.42}{}$, raw data in grey.
}
\end{figure}
We continue to present a thorough characterization of a single (typical) quantum emitter.
Figure~\ref{fig3}(a) depicts a confocally recorded low-temperature PL spectrum at the end of a \SI{3}{\micro\meter} long plasmonic waveguide excited by \SI{2}{\micro\watt} of HeNe laser power.
The spectrum exhibits a single sharp and spectrally well-isolated emission line and some residual defect PL, indicating the presence of just a single emitter at this position.
The inset shows the spectral line shape of this emitter and a Lorentzian fit to the data, from which we extract the time integrated linewidth to be \SI{2.64 \pm 0.02}{\milli\electronvolt}.
Figure~\ref{fig3}(b) shows polarization dependent PL intensity and a fit to the data using Maulus' law for comparison to the statistical data presented in Fig.~\ref{fig2}(d).
We extract the degree of linear polarization $\frac{I_{max}-I_{min}}{I_{max}+I_{min}}$ of \SI{85.7}{\percent} and clearly observe that the principal polarization axis is oriented along V, i.e., perpendicular to the long waveguide axis, as indicated in orange in the same panel.
This is consistent with the enhanced polarization ratio discussed in Fig.~\ref{fig2}(d), indicating that this particular emitter has formed due to the topography provided by the underlying waveguide.
To perform time-resolved PL measurements we employed a pulsed laser diode sending \SI{1}{\micro\watt} CW-equivalent power onto the sample with a pulse duration of \SI{\sim 90}{\pico\second}, at a repetition rate of \SI{40}{\mega\hertz}.
The resulting time-resolved PL intensity is plotted in Fig.~\ref{fig3}(c) on a semi-logarithmic scale.
A bi-exponential decay fit to the raw data reveals the dominant emitter lifetime of \SI{1.07 \pm 0.01}{\nano\second} and a second weaker contribution with a lifetime of \SI{3.5 \pm 0.1}{\nano\second}.
To test the photon statistics we performed a second-order correlation function measurement using a Hanbury-Brown and Twiss setup with confocal detection shown in Fig.~\ref{fig3}(d).
As is clearly visible from the data, the peak for zero delay time shows a significantly reduced number of correlations, indicating sub-Poissonian photon statistics consistent with the model that the emission line is due to a single emitter.
To enhance visibility, the individual peaks are binned and displayed as a histogram in black. Subsequent comparison of the peak areas results in a $g^{(2)}(0) = 0.42$, proving the single photon nature of emission from this QD.

\begin{figure}[!ht]
\scalebox{\figurescale}{\includegraphics{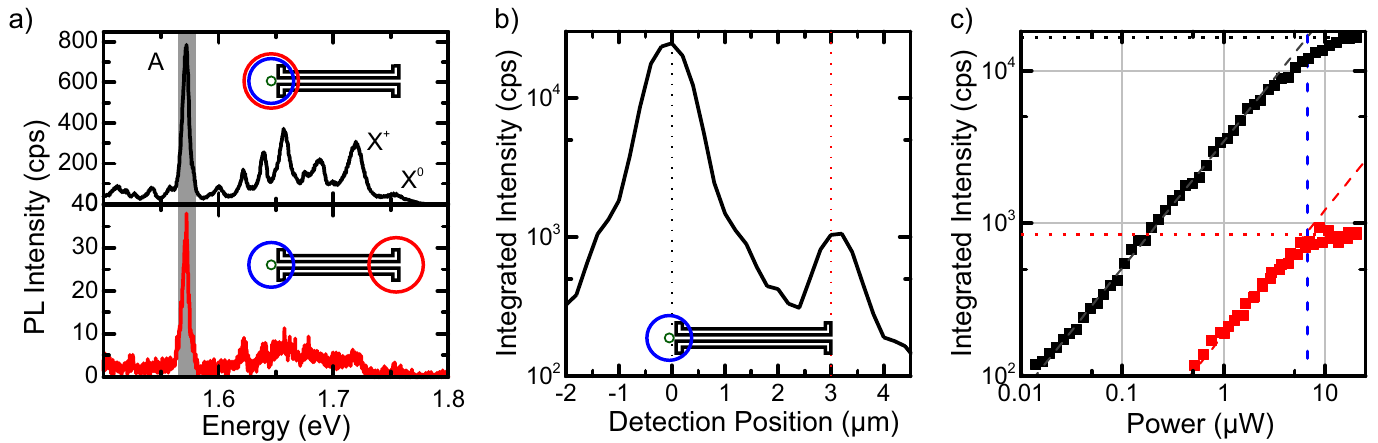}}
\renewcommand{\figurename}{Figure}
\caption{\label{fig4}
Coupling between single emitter and plasmonic slot waveguide.
(a) Confocal PL spectrum recorded at the QD position (black curve). The red spectrum recorded with excitation centered on the QD position, detection located at the remote end of the plasmonic slot waveguide.
(b) Detected PL signal integrated over highlighted area in (a) as a function of detection position along the waveguide, zero indicating confocal measurement. Initial spatial decay of PL intensity in agreement with detection spot size, increase of PL intensity at \SI{3}{\micro\meter} consistent with identical waveguide length.
(c) Power dependent measurement for confocal and plasmon-mediated configuration in black and red, respectively.
}
\end{figure}

We continue to explore the interaction between single emitters and the plasmonic waveguides.
The top panel in Fig.~\ref{fig4}(a) depicts a low-temperature PL spectrum recorded in the confocal geometry, located at the end of a \SI{3}{\micro\meter} long waveguide.
The spectrum shows PL intensity from X$^0$ and $X^+$, as well as several sharp emissions lines visible at lower energy.
Here we concentrate on the brightest emission line, labelled A on the figure, centered at \SI{1.572}{\electronvolt}.  By fixing the excitation laser to the emitter position and moving the detection position to the remote end of the waveguide, we recorded the non-confocal PL intensity depicted in the lower panel in Fig.~\ref{fig4}(a).  Here, the measured intensity is greatly reduced due to the spatial separation between excitation and detection positions, yet, the remaining sharp emission line at \SI{1.572}{\electronvolt} shows an identical spectral footprint to the confocal spectrum.  This observation indicates that the QD located at the other end of the waveguide couples directly to the plasmonic mode of the waveguide.  To prove this expectation, Fig.~\ref{fig4}(b) shows the spectrally integrated PL intensity within the shaded region denoted on panel (a) as a function of detection position along the waveguide axis with the excitation position fixed to the emitter location.
Position zero corresponds to the confocal measurement geometry exhibiting the highest PL intensity.
When moving the detection position away from the confocal geometry, the PL intensity decreases with a spatial Gaussian decay length of \SI{1.52 \pm 0.02}{\micro\meter} consistent with the detection spot size of \SI{1.58}{\micro\meter}.  However, at a separation of \SI{3}{\micro\meter}, a significant increase in PL intensity is observed consistent with scattering of plasmons into the far-field by the out-couplers at the end of the \SI{3}{\micro\meter} long plasmonic waveguide.
Due to the spectrally identical PL signature and the re-appearance of the PL signal at the far end of the waveguide, we conclude that the emitter located at one end of the waveguide spontaneously emits into far-field modes as well as into propagating surface plasmon polaritons guided by the plasmonic waveguide.
To gain insight into the coupling efficiency, Fig.~\ref{fig4}(c) shows power-dependent measurements performed at both ends of the waveguide - detected either confocally (black curve) or from the remote end of the plasmonic waveguide (red curve).
For low-power HeNe laser excitation, the intensity of both datasets scales with the incident power with an exponent of \SI{0.83 \pm 0.01}{} and \SI{0.81 \pm 0.02}{}, respectively, and both saturate at higher excitation powers as expected for single photon emitters, for excitation powers higher than \SI{6.7}{\micro\watt}.
This agreement of both quantities strongly indicates that both measurements address the same quantum emitter.
In the confocal geometry, the saturation count rate of \SI{1.65}{\kilo\hertz} is $19.5 \times$ the saturation count rate of \SI{0.85}{\kilo\hertz} in non-confocal detection.

\begin{figure}[!ht]
\scalebox{\figurescale}{\includegraphics{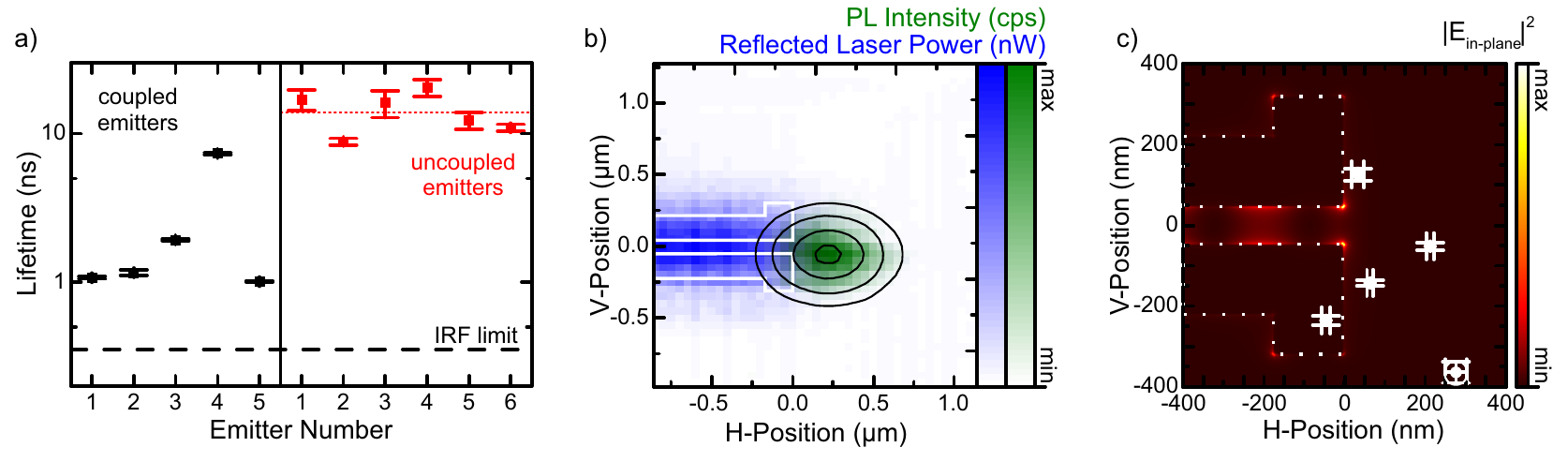}}
\renewcommand{\figurename}{Figure}
\caption{\label{fig5}
Lifetime and location statistics on emitters coupled to plasmonic waveguides
(a) Decay lifetime measurements on coupled (uncoupled) emitters in black (red), red dotted line indicating average lifetime of \SI{14 \pm 3}{\nano\second} for uncoupled emitters. Black dashed line indicates IRF time resolution limit.
(b) Concurrent high-resolution PL (green) and laser reflectivity (blue) scan revealing exact relative position of waveguide and QD A, discussed in Fig.~\ref{fig4}. QD position fitted by two-dimensional Gaussian fit and waveguide extracted from SEM data outlined in black and white, respectively.
(c) Simulation of the distribution of $|E_{in-plane}|^2$ for excited plasmonic waveguide end proportional to the plasmonic coupling efficiency. Location of coupled emitters from panel (a) indicated in white.
}
\end{figure}

In principle, the ratio of the saturation intensities for the measurements presented in Fig.~\ref{fig4}(c) can be used as a measure for the extraction efficiency of the emitter PL through the waveguide.
However, both the far-field radiation pattern and the coupling efficiency to the waveguide are exceptionally sensitive to the precise position of the emitter with respect to the plasmonic waveguide.
Therefore, the coupling between the emitters and the corresponding waveguides is assessed via the effective Purcell factor $F_P = \frac{\Gamma _{coupled}}{\Gamma _{uncoupled}}$ by measuring the decay lifetime of the emitters.
Since the emitters examined in this manuscript are induced by the underlying topography of the waveguide, and, thus, cannot be investigated emitting purely into vacuum photonic modes, their lifetime is compared to the average lifetime of several ($N=6$) emitters located far away from a plasmonic waveguide.
In Fig.~\ref{fig5}(a), we present the decay times of 5 (6) coupled (reference) dots plotted with the black (red) data points.
The average lifetime of the reference quantum dots is found to be \SI{14 \pm 3}{\nano\second}.
The Purcell factors of the five waveguide coupled emitters shown in Fig.~\ref{fig5}(a), relative to the average decay rate for uncoupled centers, are calculated to vary between \SI{1.9 \pm 0.4}{} and \SI{15 \pm 3}{}, where the emitter presented in Fig.~\ref{fig4} exhibits a lifetime of \SI{7.4 \pm 0.1}{\nano\second} corresponding to a Purcell factor of $F_P = \SI{1.9 \pm 0.4}{}$.
Emitter lifetimes reported in the literature for localized excitons in WSe$_2$ vary strongly from study to study\cite{tripathi.2018} depending on the nature of the emission center and details of the local exciton confinement potential.

To elucidate the impact of the positioning of emitters relative to the waveguide on the coupling efficiency, we determined the position of the emitter analyzed in Fig.~\ref{fig4} with a precision \SI{< 50}{\nano\meter} by recording its point spread function and that of the waveguide via laser reflectivity.
The green color map in Fig.~\ref{fig5}(b) shows the result using a step size of only \SI{50}{\nano\meter}.
We simultaneously monitor the reflected laser power illustrated by the blue colormap and by fitting extract the position of the waveguide (white curve).
The black contour lines indicate the results of a Gaussian fit to the PL intensity data, determining the most likely emitter position relative to the waveguide end $(H, V) = (0,0)$ to be $\Delta H = \SI{210 \pm 10}{\nano\meter}$ and $\Delta V = \SI{-50 \pm 10}{\nano\meter}$.
Further details on the fitting procedures are provided in Supplementary Materials.
Clearly, the emitter is fortuitously positioned relative to the waveguide, such as to allow coupling into the plasmonic mode (Fig.~\ref{fig4}), but it is also not maximally overlapping with the local plasmonic field, thus accounting for the relatively low measured Purcell factor of \SI{1.9 \pm 0.4}{}.

Figure~\ref{fig5}(c) shows the extracted relative positions for the other coupled emitters introduced in panel (a), the error bars indicate the $2\sigma$ Gaussian fit error of the positioning. The waveguide location is indicated by the white dotted line.
The false color data depicts the in-plane electric field intensity, expected to be proportional to the emission rate into the plasmonic mode for polarization-averaged emitters.
This field distribution varies strongly over length scales of only a few nm and the emitters are distributed around the out-coupling structure.
Consequently, the calculated normalized coupling efficiencies for the marked emitters are distributed between $8 \times 10^{-5}$ and $0.006$.
In our measurement, this is reflected by a large spread in the lifetime from \SIrange{1}{7}{\nano\second}.
Furthermore, it is apparent that none of the measured emitters is located at an absolutely optimum location to produce maximum coupling efficiency.
Even the highest of the calculated emission rates is a factor of $\sim 160$ smaller than the corresponding rate for the optimum position, indicating that the measured Purcell factors of up to $15 \pm 3$ could still increase further.
The best-case scenario could be approached by exerting more control during strain engineering to ensure that emitters are preferentially created closer to the optimum positions.
Furthermore, additional enhancements of the plasmonic coupling could be achieved by improving the emitter quality, e.g. by encapsulation in hexagon boron nitride which has been demonstrated to reduce non-radiative processes for free\cite{ajayi.2017,cadiz.2017,wierzbowski.2017} and bound excitons\cite{Tonndorf.2015}.

\subsection{Summary}
In summary, we have characterized the emissive properties of an atomically thin layer of WSe$_2$ exfoliated on top of a plasmonic slot waveguide.
PL measurements performed at cryogenic temperature revealed the presence of strongly localized excitons with emission linewidths between \SI{2}{\milli\electronvolt} and \SI{20}{\milli\electronvolt}, mostly concentrated around the outcoupling structure of the waveguide.
The density of emitters was found to be increased by a factor of at least $(3.3 \pm 0.7) \times$ at positions where the strain field in the flake is high due to the topography of the underlying waveguide structure.
Single-photon emission from these emitters is demonstrated by autocorrelation measurements yielding $g^{(2)}(0) = 0.42$.
Finally, the observation of identical spectral features and power dependence of luminescence at both ends of the waveguide confirms coupling to the plasmonic mode.
This hybrid nano-photonic device is, thus, capable of generating and routing single photons and plasmons at the nanoscale.
Lifetime measurements show Purcell factors between \SI{1.9 \pm 0.4}{} and \SI{15 \pm 3}{} and numerical assessment of the theoretical coupling rates indicates strong potential for further optimization by strain engineering.

\begin{acknowledgement}
We gratefully acknowledge financial support from the DFG via the German Excellence Initiative via NIM, as well as support of the Technische Universit\"at M\"unchen (TUM) - Institute for Advanced Study, funded by the German Excellence Initiative and the TUM International Graduate School of Science and Engineering (IGSSE).
J.C. is supported by NSF-DMR1410599 and the Visiting Professor Program from the Bavarian State Ministry for Science, Research \& the Arts.
\end{acknowledgement}

\section{Author contributions statement}
M.K., J.J.F and M.B. designed the study. O.H. designed and fabricated the waveguide structures, M.P. exfoliated and transferred monolayers. M.B. built the optical setup and together with M.J. conducted optical measurements and performed the data analysis with support by J.C. and G.V.; O.H., M.J. and M.B. implemented FDE and FDTD simulations.
All authors discussed the results. M.B. and G.V. wrote the manuscript with contributions from all other authors.
J.J.F. and M.K. inspired and supervised the project.

\bibliography{paper}

\end{document}